\documentstyle[12pt,aaspp]{article}
\def\half{\textstyle {1 \over 2} \displaystyle}
\def\third{\textstyle {1 \over 3} \displaystyle}
\def\ltsima{$\;\buildrel < \over \sim \;$}
\def\simlt{\lower.5ex \hbox{\ltsima}}
\def\gtsima{$\;\buildrel > \over \sim \;$}
\def\simgt{\lower.5ex \hbox{\gtsima}}
\begin{document}
\title{ Far-Infrared Water Line Emissions From \\
Circumstellar Outflows}
\author{Wesley Chen and David A. Neufeld}
\affil{Department of Physics \& Astronomy,
The Johns Hopkins University, Baltimore, MD 21218}

\begin{abstract}
We have modeled the far-infrared water line emission expected
from circumstellar outflows from oxygen-rich late-type stars, as a
function of the mass-loss rate and the terminal outflow velocity.  For
each mass-loss rate and terminal outflow velocity considered, we
computed self-consistently the gas density, temperature, outflow
velocity, and water abundance as a function of distance from the star.
We then used an escape probability method to solve for the equilibrium
level populations of 80 rotational states of water and thereby obtained
predictions for the luminosity of a large number of far-infrared rotational
transitions of water.  In common with previous models, our model predicts
that water will be copiously produced in the warm circumstellar gas, and that
water rotational emission will dominate the
radiative cooling.   {\it However, our use of a realistic radiative cooling
function for water leads to a lower gas temperature than that
predicted in previous models.  Our predictions
for the far-infrared water line luminosities are consequently significantly
smaller than those obtained in previous studies.}  Observations to be carried
out by the Infrared
Space Observatory will provide a crucial test of the models presented here.
\end{abstract}


\section{INTRODUCTION}

Theoretical models (e.g. Goldreich \& Scoville 1976, hereafter GS76)
for the physical and chemical conditions in circumstellar outflows
make the strong prediction that water will be copiously produced in
oxygen-rich outflows; water is expected to account
for almost all of the gas-phase oxygen that is not bound as CO, and
far-infrared water emissions are expected to dominate the radiative cooling of
the gas (Neufeld \& Kaufman 1993, hereafter NK93).
This prediction is supported by observations of
luminous water maser emissions from hundreds of such outflows
(Bowers \& Hagen 1984 and references therein), with
luminosities that are broadly
consistent with large water abundances (Cooke \& Elitzur 1985).

Although several submillimeter water maser transitions have been
detected from oxygen-rich circumstellar outflows (Menten et al.~1990a,b;
Melnick et al.~1993), none of
the far-infrared non-masing transitions that dominate the cooling
of the gas has yet been observed, because absorption by the Earth's
atmosphere makes such observations infeasible even at airplane
altitude.   However, the Infrared Space Observatory (ISO) satellite,
scheduled for launch in 1995, should provide an ideal instrument
for the study of far-infrared cooling transitions of water molecules
within warm astrophysical gas.

In this {\it Letter}, we present new predictions for the expected
far-infrared water line luminosities from oxygen-rich circumstellar
outflows.  In \S2, we describe our
treatment of the physical and chemical processes that determine
the dynamics, temperature,
chemical composition, and water line emissivities within such outflows.
In \S3 we present our results for the structure of the outflowing
gas and for the far-infrared water emission line strengths.
In \S4, we discuss the significant differences between the results obtained
here and those obtained in previous studies.

\section{MODEL DESCRIPTION}
\subsection{Dynamics}
Assuming that the circumstellar outflow is accelerated primarily
by radiation pressure upon newly-formed dust grains (Gilman 1972;
Salpeter 1974; Kwok 1975),
we may describe the outflow velocity of the gas by the equation
\begin{equation}
v\frac{dv}{dr}=\frac{\kappa(r)L}{4\pi r^2c}-\frac{GM}{r^2},
\end{equation}
(GS76), where $v(r)$ is the velocity of the gas at distance $r$
from the center of the star,
$\kappa(r)$ is
the dust opacity, $L$ is the stellar luminosity, $c$ is the speed of light,
and $M$ is the mass of the star.

For simplicity, we followed GS76 in assuming that the dust opacity
increases steadily with $r$ according to the expression
\begin{equation}
\kappa(r)=\frac{Qn_d\sigma_d}{\rho}
=\Gamma\left(1+\frac{\Delta r^2}{[(10R)^2+r^2]}\right),
\end{equation}
where $R$ is the stellar radius, $\Gamma$ and $\Delta$ are constants that
are determined by the initial and terminal gas outflow velocities,
$n_d$ and $\sigma_d$ are the number density and typical
geometrical cross section of the dust grains, and $Q$ is the typical
ratio of the infrared absorption cross section to the geometrical
grain cross section.  We take $0.03$ as an effective value for $Q$,
adopting the
results computed by Justtanont et al. (1994; hereafter JST94)
for silicate grains with a power-law size distribution (Mathis, Rumpl, \&
Nordsieck 1977).  Given a particular value of the terminal outflow velocity,
the exact dependence of $\kappa$ upon $r$ does not greatly affect
the predicted strengths of the water lines considered in this
{\it Letter}, because we have confined our attention to transitions
that are primarily excited outside the region where the
gas and dust are accelerated.  Thus although more
detailed models can be constructed for
the variation of the opacity with position (e.g. JST94), the effect upon the
predicted line strengths is small for the transitions considered
here.

\subsection{Thermal Balance}
The gas temperature $T$ in the outflow satisfies the equation
\begin{equation}
\frac{r}{T}\frac{dT}{dr}=-\frac{4}{3}\left(1+\frac{1}{2}\frac{d{\rm ln}v}
{d{\rm ln}r}\right) + \frac{2r(H-C)}{3vn_{\rm H}kT(x_H+x_{H2}
+x_{He})}
\end{equation}
(GS76), where $n_{\rm H}$ is the density of hydrogen nuclei, and $x_H$,
$x_{H2}$, and $x_{He}$ are respectively the abundances of H,
${\rm H}_2$, and He relative to hydrogen nuclei.
Again following GS76, we
assume $x_H=x_{H2}=\third$ and $x_{He}=0.1$, although
the results depend only very weakly upon the exact values adopted.
The first term on the
right hand side describes the adiabatic cooling rate, and $H$ and $C$ are the
heating and cooling rates per unit volume due to other processes.

The additional heating and cooling processes we considered are grain-gas
collisional
heating, H$_2$O rovibrational heating (resulting from the absorption
of stellar radiation by water molecules),
and H$_2$O rovibrational cooling.  We have also considered CO and
H$_2$ cooling and found them to be negligible.
The grain-gas collisional heating
rate is $\half \rho v_d^3\sigma_d n_d$,
where $v_d$
is the drift velocity of the dust relative to the gas.
Provided $v_d$ is much greater than
the sound speed in the gas, the drift velocity is
given by $v_d=(QLv/{\dot M}c)^{1/2}$, where
${\dot M}$ is the mass loss rate
(GS76; Tielens 1983).  For
H$_2$O rovibrational cooling, we used the cooling functions
of NK93; the cooling rate
is a function of the temperature, the gas density, and an optical depth
parameter which depends on the water density and
the velocity gradient in the outflow.   We estimated the
{\it heating} rate due to the absorption of stellar radiation by water
molecules
as
\begin{equation}
H_{\rm H2O} = W C_{\rm H2O}(T_*),
\end{equation}
where $C_{\rm H2O}(T)$ is the water {\it cooling} rate given by NK93 for gas
temperature $T$, where $4 \pi W= 2\pi (1-[1-(R/r)^2]^{1/2})$ is the
solid angle subtended by the star, and $T_*$ is the temperature of
the stellar photosphere, which we assume to radiate as a blackbody.
This approximate expression shows the correct behavior $H_{\rm H2O}=C_{\rm
H2O}$
required by thermodynamics in the limit where the gas is surrounded
($W=1$) by radiation with a temperature equal to the gas temperature.
Our present treatment of the thermal balance neglects the effects of
heating that may result from the absorption of dust continuum
radiation by water; based upon theoretical models for the dust continuum
radiation (S. Doty 1995, private communication),
we expect such effects to be important only close
to the star, i.e. inside the region that gives rise to the water
line emission considered in this paper.
\subsection{Chemistry}
For the purposes of this study, we needed only to determine the
abundance of water.  Accordingly, we
based our computation of the abundances of the major oxygen bearing species
upon a very limited chemical network.
At temperatures $\simgt 300$~K, water is formed rapidly by
the reaction sequence:
\[ {\rm O + H_2 \rightleftharpoons OH + H} \]
\[ {\rm OH + H_2 \rightleftharpoons H_2O + H}  \]
The rate coefficients for these reactions have been given
by Wagner and Graff (1987).
Due to the endothermic nature of the first forward reaction, and
the relatively high activation barrier of the second forward reaction,
this reaction sequence is effective only in the warm ($T \simgt 300$~K)
regions close to the star.   Far away from the
star, the H$_2$O and OH molecules are photodissociated by the interstellar
radiation field:
\[ {\rm H_2O +  h\nu} \rightarrow {\rm H + OH} \]
\[ {\rm OH +  h\nu} \rightarrow {\rm O + H}. \]
The photodissociation rates for OH and H$_2$O have been calculated by
Roberge et al. (1991) as
a function of visual extinction within a plane-parallel slab;
we have modified their results as appropriate for the spherically-symmetric
geometry of relevance to circumstellar outflows.
The rate equations for all these chemical processes were integrated
along with equations (1) and (3) to yield the temperature, density,
outflow velocity and chemical abundances as a function of $r$.

\subsection{H$_2$O line emission}
Once the temperature, density, velocity, and abundance profiles
were computed, we used an escape probability method (NK93)
to obtain
the rotational level populations of water and the water line emissivities
as a function of $r$.
Integrating the line emissivities over the entire outflow region,
we obtained the luminosity for each water transition.

Our present treatment of the water line emission
neglects the effects of radiative pumping by dust continuum
radiation; as in \S2.2, we expect such effects
to be unimportant in the region that gives rise to the water
line emission considered in this paper.  We defer to a future paper
a more complete analysis in which the line radiation, dust continuum
radiation and thermal balance are treated entirely self-consistently
by an iterative method; such a calculation will be needed to
obtain reliable predictions for water transitions of
higher excitation than those considered here.

\section{RESULTS}
Figures 1, 2, and 3 describe the structure of the outflow
region for the same set of parameters considered by GS76:
stellar mass $M=2\times 10^{33}$~g;
stellar luminosity $L=4\times 10^{37}$ erg s$^{-1}$;
stellar radius $R=6\times 10^{13}$~cm;
stellar temperature $T_{*}=
2000$ K;  mass outflow rate ${\dot M}=3\times 10^{-5}M_{\odot} \rm yr^{-1}$;
terminal outflow velocity $v_{\infty}=20$ km s$^{-1}$; and
initial oxygen abundance $n({\rm O})/n_{\rm H}=5\times 10^{-4}$.

Figure 1 shows the strength of
various cooling and heating processes as a function of the distance $r$ from
the star.  The heating is dominated by the dust-gas collision
process except close to the star where water vibrational heating is important.
The water rotational cooling rate drops rapidly between $r=10^{16}$ and
$r=10^{17}$~cm due to
the photodissociation of H$_2$O molecules by the interstellar UV field.  The
effects of photodissociation are also apparent in Figure 2,
which shows the abundances of O, OH and H$_2$O.
For $r < 3\times 10^{16}$~cm, photodissociation is negligible and
H$_2$O accounts for almost all the gas-phase oxygen that is
not bound as CO.   The resultant temperature profile is
shown in Figure 3, together with the analogous result from GS76.

Example line strengths are given in Table 1 for three different
mass-loss rates.  Here we compare
the far-infrared line strengths predicted by our model with
those obtained previously by Deguchi and Rieu (1990; hereafter DR90),
{\it who assumed the GS76 temperature profile in their calculation.}
These results apply to a the same terminal outflow velocity
($v_\infty=10$~km~s$^{-1}$) and the same initial
oxygen abundance ($n({\rm O})/n({\rm H_2})=4 \times 10^{-4}$) assumed by DR90
for the case ${\dot M}= 10^{-5}M_{\odot} \rm yr^{-1}$.
Table 1 lists the predicted fluxes for seven transitions of ortho-water
for which ISO observations are planned, given an assumed distance
to the star of 100~pc.  The
final  column in Table 1 shows the ratio of our calculated fluxes for
${\dot M}=10^{-5}M_{\odot} \rm yr^{-1}$ to those obtained by DR90.
As expected given the lower gas temperatures predicted in the present
study (see Figure 3), our predicted line strengths are smaller than
those of DR90 by substantial factors (ranging from 7 to 24).

In Figure 4, we examine the dependence of the far-infrared water
line luminosities upon the assumed abundance of gas-phase oxygen not
bound as CO (or, equivalently, upon the assumed water abundance).
The results shown here apply to the outflow from the star IRC+10011,
given the following assumed parameters (Knapp \& Morris 1985, JST94):
stellar mass $M_*=1\,M_{\odot}$, mass loss rate
${\dot M}=1\times10^{-5}M_{\odot}\rm yr^{-1}$, stellar temperature
$T_{*}=2000K$,
stellar radius $R=4.5\times 10^{13}$~cm, terminal outflow velocity
$v_{\infty}=23$~km s$^{-1}$, and distance to the star = 480~pc.
For most of the emission line strengths shown in Figure 4, there is no
strong dependence upon the water abundance, because water is
the dominant coolant.  Thus an increase in the
assumed water abundance leads to an
decrease in the gas temperature so that the net rate of water emission
remains unchanged.  Furthermore, much of the emission is generated within
regions where the lines are optically thick.

\section{DISCUSSION}
Our predicted line strengths for far-infrared water emissions
from oxygen-rich circumstellar outflows are substantially smaller
than those obtained previously by DR90.  This discrepancy arises because
our model predicts a gas temperature in the outflow which is
much smaller than that obtained by GS76.  The lower temperature
obtained in the present study has two origins.
First, as noted by JST94, GS76 adopted an unrealistically large
value of 0.5 for $Q$, the typical
ratio of the infrared absorption cross section to the geometrical grain
cross section.
This in turn led to overestimates of the grain drift velocity $v_d$
and of the gas-grain collisional heating rate.  Using the more realistic
value $Q=0.03$, which we adopt in the present study, JST94 obtained
a gas temperature in the outflow which was considerably smaller than
that of GS76.  Second, the simple three-level model for the water molecule
adopted by both GS76 and JST94 leads to an estimate of the water cooling
rate which lies significantly {\it below} the more realistic cooling
function derived by NK93.

Observations of water maser emission from circumstellar outflows
may provide additional support for the lower outflow temperatures
that we derive.  Theoretical models for such emission, constructed
by Cooke \& Elitzur (1985) (and subsequently by Neufeld \& Melnick 1991)
{\it on the basis of the GS76 temperature profile},
predicted maser luminosities for high mass-loss rate outflows that
significantly
exceed what is observed in any real source.  Cooke \& Elitzur (1985)
suggested that their neglect of photodissociation might have led
to an overestimate of the maser luminosity.   We regard that
explanation as unlikely, since for mass loss rates in excess of
${\dot M}= 3 \times 10^{-5}M_{\odot} \rm yr^{-1}$, water is significantly
photodissociated only at radii $\simgt 10^{16}$~cm  whereas the
maser emission is generated within a region of radius $\sim 10^{15}$~cm. We
speculate instead that previous theoretical
models overestimated the maser line luminosity because
they were based upon an
overestimate of the gas temperature in the outflowing gas.  The inclusion
of maser transitions in our model will be needed to test this
speculation.

We thank S. Doty for providing us with estimates for the dust
continuum radiation field in IRC+10011,
and we gratefully acknowledge the support of NASA grant NAGW-3147
from the Long Term Space Astrophysics Research Program and of
NASA grant NAGW-3183.

\newpage

\begin{table}[h,b,t]
	\begin{center}
	\caption{Predicted Emission Line Fluxes and Comparison to DR90
Results}
	\vspace{0.3cm}
  	\begin{tabular}{l|r|l|l|l|l} \hline\hline
	Transition  & Wavelength \hfill &
	\multicolumn{3}{c}{Line Flux (W cm$^{-2}$)$^{a}$} &
	CN/DR ratio$^{b}$ \\

	& ($\mu$m) \hfill &
	${\dot M}=10^{-4}$ & ${\dot M}=10^{-5}$ & ${\dot M}=10^{-6}$ &
	for ${\dot M}=10^{-5}$ \\ \hline

 $2_{  1  2 }- 1_{  0  1}$ &   179.5265 & 2.1$\times 10^{-19}$ &
1.9$\times 10^{-19}$ & 1.2$\times 10^{-19}$ & 0.071 \\
 $3_{  3  0 }- 3_{  2  1}$ &   136.4944 & 2.0$\times 10^{-19}$ &
5.9$\times 10^{-20}$ & 1.7$\times 10^{-20}$ &       ---- $^c$ \\
 $4_{  1  4 }- 3_{  0  3}$ &   113.5366 & 3.6$\times 10^{-19}$ &
1.6$\times 10^{-19}$ & 9.5$\times 10^{-20}$ & 0.043 \\
 $2_{  2  1 }- 1_{  1  0}$ &   108.0730 & 4.3$\times 10^{-19}$ &
2.3$\times 10^{-19}$ & 1.2$\times 10^{-19}$ & 0.052 \\
 $2_{  2  0 }- 1_{  1  1}$ &   100.983  & 4.7$\times 10^{-19}$ &
2.1$\times 10^{-19}$ & 6.8$\times 10^{-20}$ & 0.15  \\
 $5_{  0  5 }- 4_{  1  4}$ &    99.4926 & 4.3$\times 10^{-19}$ &
1.6$\times 10^{-19}$ & 7.8$\times 10^{-20}$ & 0.041 \\
 $5_{  1  5 }- 4_{  0  4}$ &   95.6261  & 4.2$\times 10^{-19}$ &
1.4$\times 10^{-19}$ & 5.1$\times 10^{-20}$ & 0.11  \\
 $6_{  1  6 }- 5_{  0  5}$ &    82.0304 & 5.0$\times 10^{-19}$ &
1.7$\times 10^{-19}$ & 7.3$\times 10^{-20}$ & 0.055 \\
	\hline\hline
	\end{tabular}
	\end{center}
\end{table}
\noindent $^{a}$ Line flux for a terminal outflow velocity $v_{\infty}=10
\rm \,km\,s^{-1}$, a
water abundance $n($H$_2$O)/$n($H$_2)=4\times 10^{-4}$, and an assumed
distance to the source of 100~pc.

\noindent $^{b}$ Ratio of the results of this study
(CN) to those of DR90 (DR).

\noindent $^{c}$ Transition not listed by DR90

\newpage

\section*{FIGURE CAPTIONS}

\noindent FIGURE 1:  The heating and cooling rates
for various processes, as a
function of distance $r$ from the star, for
the set of outflow parameters adopted by GS76 (see text).

\noindent {FIGURE 2:}  The abundances of O, OH and H$_2$O
relative to hydrogen nuclei, as a function
of distance from the star, for
the set of outflow parameters adopted by GS76 (see text).

\noindent {FIGURE 3:}  Gas temperature, as a
function of distance $r$ from the star, for
the set of outflow parameters adopted by GS76 (see text).
The GS76 temperature profile is shown for comparison.

\noindent {FIGURE 4:}  Predicted emission line fluxes for
the outflow from IRC+10011, as a function of the assumed water
abundance.  The assumed outflow parameters are given in the text.

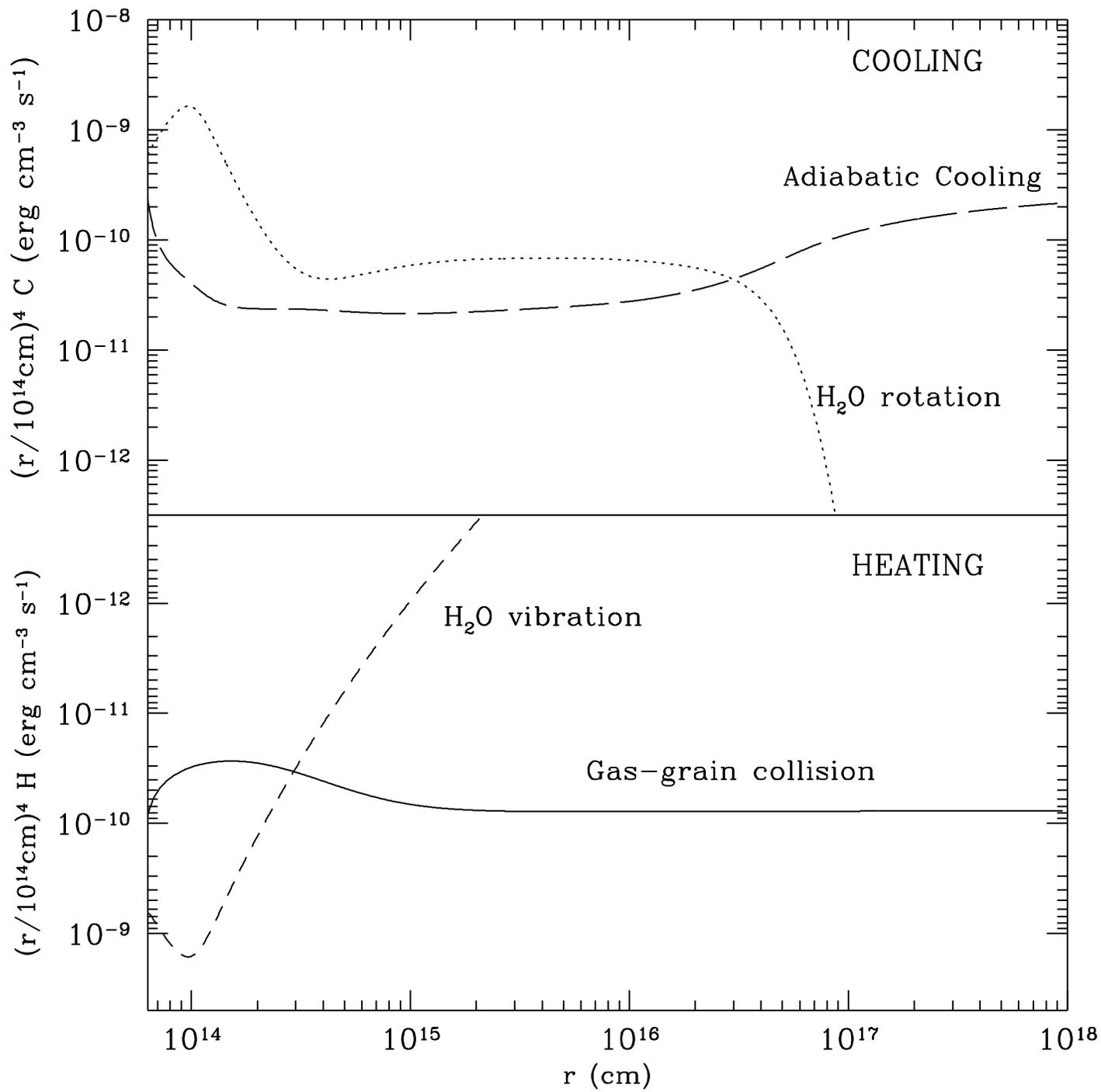
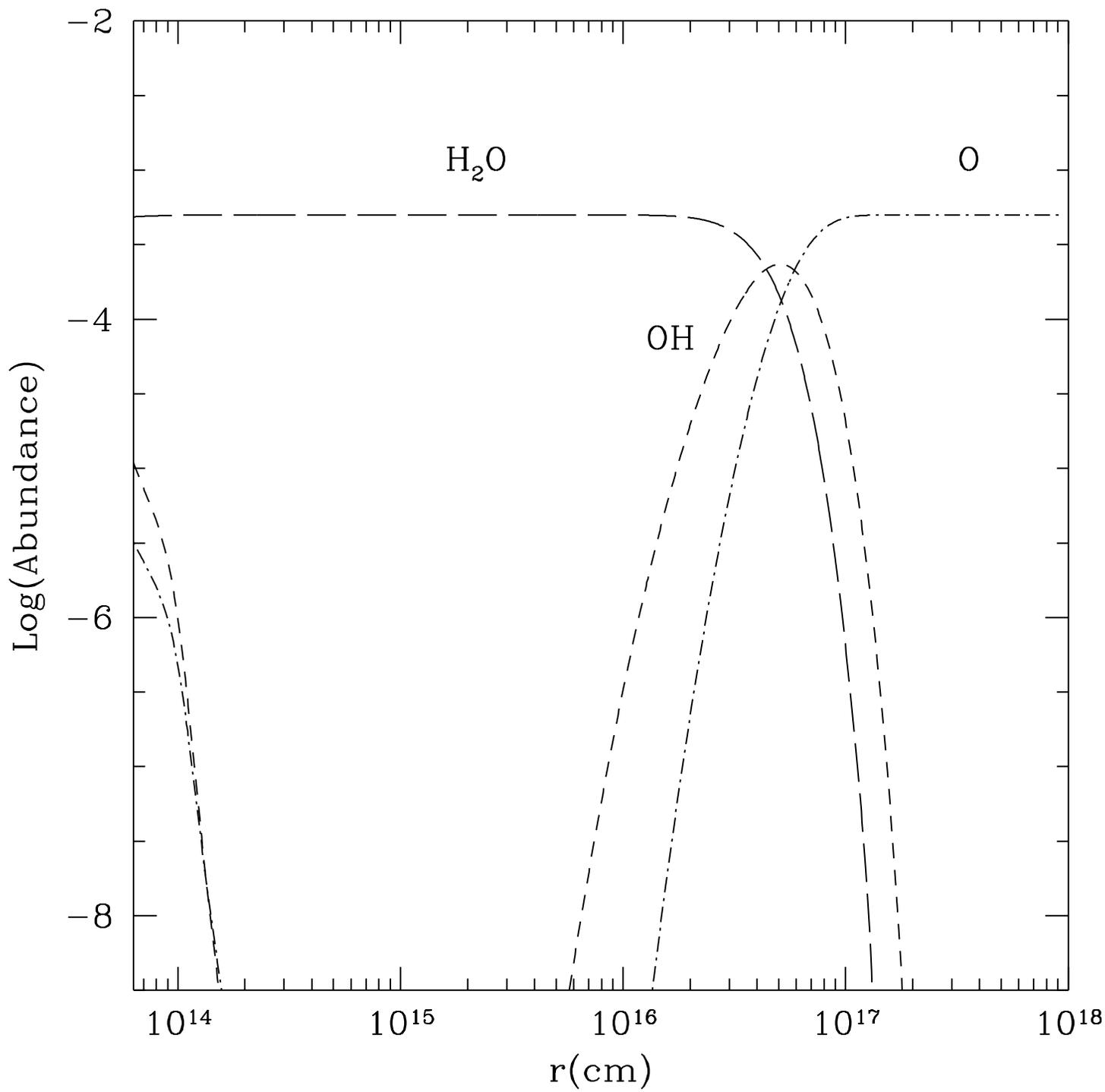
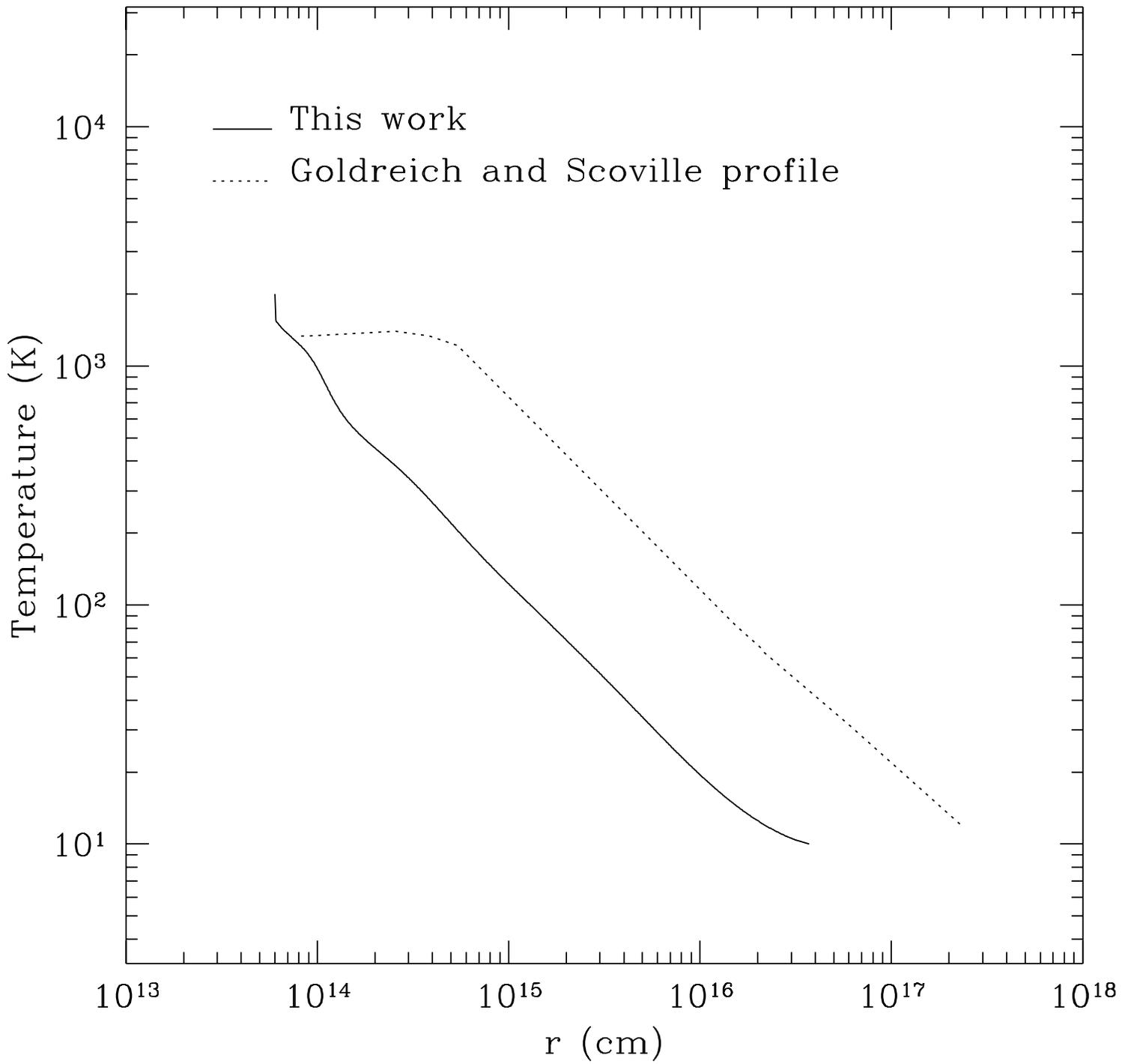
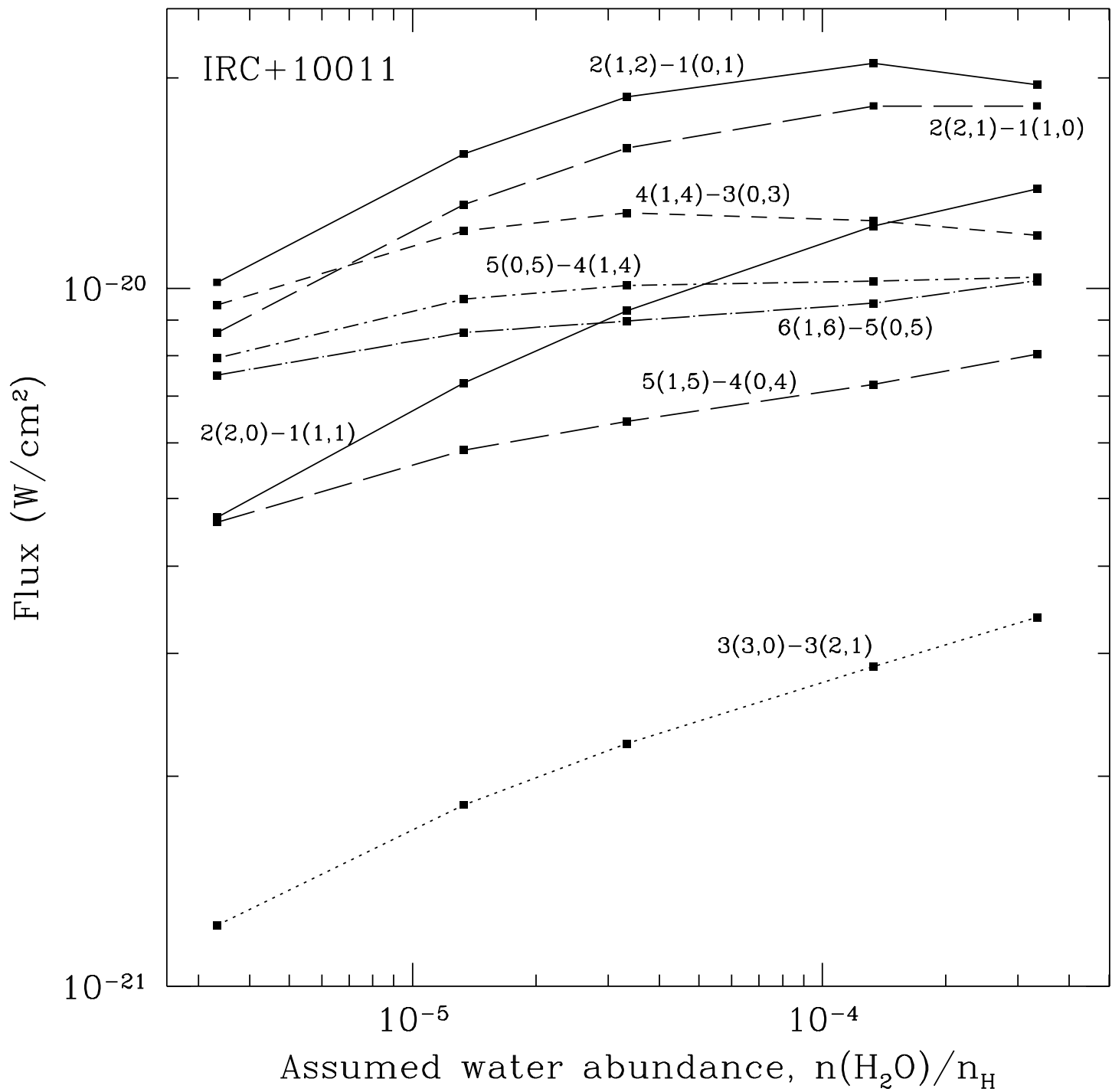
\end{document}